\documentclass[
pra,
11pt,
onecolumn,
nofootinbib,
notitlepage,
tightenlines]{revtex4-2}
\usepackage{amsmath, amsthm, mathtools, braket, amssymb, bm}
\usepackage{enumitem, subfiles, tikz, tkz-graph, tikz-cd, ifthen}
\usepackage[capitalize]{cleveref}
\usetikzlibrary{decorations.markings, quantikz}
\usepackage[raggedright, medium, explicit]{titlesec}
\usepackage{algorithm, algpseudocode}
\usepackage{float, newfloat}
\DeclareFloatingEnvironment[
    fileext= los,
    listname= {List of Algorithms},
    name= Algorithm,
    placement= tbhp
]{alg}

\titleformat{\paragraph}[runin]
{\normalfont\normalsize\bfseries}{}{0pt}{#1.}
\titlespacing{\paragraph}{0pt}
{0.5\baselineskip plus 1ex minus .2ex}{1.5ex plus .2ex}

\makeatletter

\renewcommand*{\p@subsection}{}
\makeatother

\theoremstyle{plain}
\newtheorem{thm}{Theorem}

\newtheorem{prop}[thm]{Proposition}
\newtheorem{cor}[thm]{Corollary}

\theoremstyle{plain}
\newtheorem{prob}{Problem}

\theoremstyle{definition}
\newtheorem{defn}{Definition}

\theoremstyle{remark}
\newtheorem{rmk}{Remark}[section]

\crefname{alg}{Algorithm}{Algorithms}
\crefname{prob}{Problem}{Problems}

\DeclareMathOperator{\argmax}{argmax}
\DeclareMathOperator{\plog}{polylog}
\DeclareMathOperator{\poly}{poly}

\newcommand*{\ceil}[1]{\lceil #1 \rceil}

\begin{document}

\title{The Problem of Dynamic Programming on a Quantum Computer}

\author{Pooya Ronagh\,}
\email{pooya.ronagh@uwaterloo.ca}
\affiliation{
Institute for Quantum Computing, Waterloo, ON\\
Perimeter Institute for Theoretical Physics, Waterloo, ON\\
Department of Physics and Astronomy, University of Waterloo, Waterloo, ON\\
1QB Information Technologies (1QBit), Vancouver, BC
}

\date{\today}

\begin{abstract}
We discuss the problem of finite-horizon dynamic programming (DP) on a quantum
computer. We introduce a query model for studying quantum and classical
algorithms for solving DP problems, and provide example oracle constructions
for the travelling salesperson problem, the minimum set-cover problem, and the
edit distance problem. We formulate open questions regarding quadratic quantum
speedups for DP and discuss their implications. We then prove lower bounds for
the query complexity of quantum algorithms and classical randomized algorithms
for DP, and show that no greater-than-quadratic speedup can be achieved for
solving DP problems.
\end{abstract}

\maketitle

\section{Introduction}

Solving NP-hard problems efficiently on a quantum computer has been a
challenging endeavour for quantum computation. Grover's search algorithm
\cite{grover1996fast} provides a path to achieving quadratic speedups for some
NP-hard problems such as the Boolean satisfiability problem (SAT). While this
quantum speedup is much more moderate than what is anticipated from the
exponential computing resources of quantum computers, its existence is of
theoretical and practical significance for understanding the power and
limitations of quantum algorithms. On the other hand, achieving a similar type
of speedup for other NP-hard problems, such as the travelling salesperson
\mbox{problem (TSP)}, has been a long-standing open problem for quantum
computing.

For SAT, the \emph{exponential time hypothesis} speculates that no classical
algorithm can essentially perform better than exhaustive search. However, for
TSP, the best classical algorithm is much more sophisticated than na\"ive
exhaustive search. For a TSP problem of size $n$, exhaustive search will
require enumeration of $n!$ possible solutions, whereas an algorithm of \cite
{bellman1962dynamic,held1962dynamic} based on dynamic programming (DP) solves
the problem in $O^*(2^n)$.\footnote{Here, the $O^*$ notation ignores polynomial
factors in $n$.} Perhaps this is the reason demonstrating a quadratic quantum
speedup for TSP has appeared to be difficult. Recently, \cite
{ambainis2019quantum} studied quantum algorithms for a collection of NP-hard
problems for which the best known
classical algorithms are exponentially expensive DP
solutions and showed a slightly weaker speedup (e.g., $O^*(\alpha^n)$ with the
base $\alpha$ being a number less than 2 but greater than $\sqrt 2$).
However, these results assume coherent QRAM access to solutions of smaller DP
problems and require exponentially large amounts of classical memory.

In this paper, we introduce and study the problem of dynamic programming on a
quantum computer. A DP problem is defined by a finite set
of \emph{states} $S$, a finite set of possible \emph{actions} (decisions) $A$
at each state, and a set of time epochs $\mathbb T$. Performing an action at a
given state results in a \emph{reward} (or cost) and a transition to a new
state. The goal is to find an \emph{optimal policy} for an
\emph{agent} at every state. Here, the measure of optimality is the future
reward the agent collects should it pursue the actions prescribed by a
policy. The cumulative future reward is often called the \emph
{value function}.

\paragraph{Query complexity of dynamic programming}

We consider algorithms for solving DP problems that have
query access to an oracle that,
given a state, action, and time $s \in S, a \in A$, and $t \in \mathbb T$,
simulate the effect of performing action $a$ on state $s$ at a given point $t$
in time. The goal is to solve the problem with the fewest queries to this
oracle. This model addresses the complexity of \emph{generic} algorithms for
dynamic programming, that is, those that do not possess further information
about the oracle or the particular instances of DP problems solved by them.

We focus on the case of \emph{finite horizon} DP problems, that is, those for
which $\mathbb T$ is a finite set. Finite-horizon DP problems are of great
interest in many famous applications in discrete and combinatorial algorithms,
including TSP and the minimum set-cover
problem (MSC). Despite many efforts in computer science, the best known
algorithms for these problems are exponentially expensive DP
solutions that have been known for many decades.

Finite-horizon DP problems are closely related to their infinite-horizon
counterparts. Infinite-horizon DP problems are the deterministic special cases
of Markov decision \mbox{problems (MDP)}. The latter is the subject of study in
reinforcement learning (RL). In fact, many problems solved using RL do not
possess stochastic kernels and reward functions. This includes, for instance,
classic computer games and
deterministic optimal control problems. The query complexity
of infinite-horizon DP problems is therefore of significant practical interest
in optimal control and RL. \cite{chen2017lower} proves lower bounds for
randomized classical algorithms solving MDPs using various data structures to
provide the probability distribution functions associated to the MDP transition
kernels.

\paragraph{Summary of our contributions}

In \cref{sec:dp}, we introduce a general framework for studying finite-horizon
DP problems on a quantum computer. We introduce a query model for studying
bounded-error quantum algorithms that make coherent queries to an oracle
representative of the DP problem. We also provide an analogous classical query
setting to compare such bounded-error quantum algorithms against classical
randomized algorithms. We then state several open problems
pertaining to the potential existence of quadratic quantum speedups in solving
DP problems in \cref{sec:open-quantum-problems} and discuss their implications.
In \cref{sec:examples}, we provide several example constructions for the
above-mentioned DP oracle in the case of TSP, MSC, and the edit distance
problem.

We then prove lower bounds for the query complexity of quantum algorithms and
classical randomized algorithms for solving these problems, and show that a
greater-than-quadratic speedup in the number of state--action pairs
cannot be achieved using quantum algorithms. In \cref{sec:q-lbd}, we
provide a quantum query complexity lower bound of $\Omega(\sqrt{|S||A|})$ for
solving DP problems using the generalized relational adversary method \cite
{ambainis2002quantum}. Furthermore, in \cref{sec:c-lbd}, we apply similar ideas
from adversary methods to a classical query complexity setting to prove a
lower bound of $\Omega(|S||A|)$ on the query complexity of classical
bounded-error randomized algorithms for these problems. We conclude that the
discussed potential quadratic speedups would result in optimal quantum
algorithms, up to polylogarithmic factors. In particular, we rule out the
possibility of achieving exponential quantum speedups for DP.

\section{Dynamic Programming}
\label{sec:dp}

Let $S$ and $A$ be two given finite sets of states and actions, respectively.
The actions are taken at points in a discrete set of \emph{time epochs}
$\mathbb T= \{0, \ldots, T-1\}$. In this case, the DP problem is said to have a
finite horizon, which all DP problems considered in this paper have.
The following structure is given:
\begin{enumerate}[label=(\alph*), leftmargin=16pt, rightmargin=16pt]

\setlength{\itemindent}{0.7cm}

\item Finite sets $S$ and $A$, and a transition kernel or \emph{law of motion}
$$a_t: S \to S \quad \forall t\in \mathbb T, \forall a \in A\,;$$
\item A \emph{reward structure} which is a bounded, deterministic, possibly
time-dependent function of states, actions, and time epochs, and for simplicity
takes values in the set of non-negative integers
$$r_t= r_t(s, a): S \times A \to \mathbb Z_{\geq 0}
\quad \forall\, t \in \mathbb T\,.$$
\end{enumerate}
The boundedness condition allows us to define a positive integer denoted by
$\ceil{r} \in \mathbb N$ as an upper bound on reward values. We note that,
without loss of generality (and by a constant shift of all rewards if needed),
we assume a lower bound of $0$ for the reward structure.

By the above definition of the reward structure, we have implicitly assumed
that all actions in set $A$ are \emph{admissible} for all states in $S$. For
a DP in which this condition is not naturally satisfied by the model (i.e., some
actions are not allowed at certain states), we may, without loss of generality,
let an originally inadmissible action $a$ at a state $s$ map $s$ to a null
state additionally defined.

\paragraph{Value function}

A (deterministic) policy consists of the choice of a single action at every
state and every point in time:
$$\pi_t: S \to A \quad \forall t \in \mathbb T.$$
To a policy $\pi= (\pi_t)_{\{t \in \mathbb T\}}$, we associate a possibly
time-dependent value function $V^\pi_t: S \to \mathbb Z_{\geq 0}$ defined via
$$V^\pi_t(s)= \sum_{i \geq t} r_i(s_i, a_i)\,,$$
where $s_0= s$ is an initial state, and all subsequent actions are chosen
according to the policy $\pi$. That is, $a_t= \pi_t (s_t)$ and
$s_{t+1} = a_t(s_t)$.
We assume that a marked initial state $s_0 \in S$ is given. The goal of DP
is to find an optimal policy at $s_0$ at time $t= 0$, that is, to find
\begin{equation}
\label{eq:dp-problem}
\pi^*= \argmax_\pi V_0^\pi(s_0)\,.
\end{equation}

\paragraph{Bellman's optimality criteria}

Bellman's optimality criteria~\cite{bellman1957dynamic}, for the value function
states that an optimal policy $\pi^*= (\pi_t^*)$ is associated to the (unique)
optimal value function \mbox{$V_t^*(s):= V^{\pi^*}_t(s)$} satisfying
\begin{equation}
\label{eq:func-eq}
V_t^* (s)= \max_{a} \left\{r_t(s, a) + V_{t+1}^*(a_t(s))\right\}
\quad \forall\, t \in \mathbb T\,
\end{equation}
and the boundary condition that no reward can be accumulated after the final
time epoch. That is, $V_T^*(s)= 0$ for all states $s \in S$.

It is easy to verify that if the optimal value function $(V^*_t)$ is known,
an optimal action at $s_0$ at time $t= 0$ can be found by finding the action
$a \in A $ that maximizes $r_0(s_0, a) + V_1^*(a(s_0))$.
Alternatively, an optimizer of $V^*$
in \eqref{eq:dp-problem}, that is, the $s_0$ component of an optimal policy, may
be directly calculated. Such an algorithm
can iteratively be used $O(T)$ times at subsequent
states visited by the agent until a complete optimal policy for the DP problem
has been traversed along the time horizon $\mathbb T$.

\paragraph{Query model}

We consider quantum and (randomized) classical algorithms that make queries to
the transition kernel and reward structure in order to solve a DP
problem. The quantum algorithms are considered to make coherent queries to
\begin{equation}
\label{eq:dp-coherent-oracle}
U_{\text{DP}}: \ket{s}\ket{a}\ket{t}\ket{x}\ket{y} \mapsto
\ket{s}\ket{a}\ket{t}\ket{x \oplus a_t(s)}\ket{y \oplus r_t(s, a)}.
\end{equation}
For classical algorithms, the oracle is similar but queried classically:
\begin{equation}
\label{eq:dp-classical-oracle}
O_{\text{DP}}: (s, a, t) \mapsto \left(a_t(s), r_t(s, a)\right).
\end{equation}

We note that, in many practical scenarios, any one of the transition
kernel, the reward structure, or the policies may be independent of time.
In this case they are called \emph{time homogeneous}.

Based on Bellman's recursion, we consider two algorithms for solving problem
\eqref{eq:dp-problem}. We first define the \emph{value iteration operator}
$\mathcal F^{(t)}: \mathbb Z_{\geq 0}^{|S|}
\to \mathbb Z_{\geq 0}^{|S|}$ for all $t\in \mathbb T$ via
\begin{equation}
\label{eq:classical-value-iteration}
\mathcal F^{(t)}: v_s \mapsto
\max_{a \in A} \left\{r_t(s, a) + v_{a_t(s)}\right\}
\end{equation}
and consider their recursive applications
$$v^{(T-t-1)}= \mathcal F^{(T-t-1)} (v^{(T-t)}) \quad t\in \mathbb T,$$
starting with the initial vector of all zeroes, $v^{(T)}= 0$, for all $s \in S$.
It is easy to see via induction that $v^{(T-k)}$ attains the optimal
value function at time $T-k$:
$$V^*_{T-k}(s)= \mathcal F^{(T-k)} \circ \cdots \circ \mathcal F^{(T-1)} (0).$$
Therefore, in order to find the optimal action at $s_0$ at time $t= 0$, it
suffices to find $v^{(1)}$ and find the maximizer
$$\argmax_{a \in A} \left[r_0(s_0, a) + v^{(1)}_{a_0(s_0)}\right].$$

\begin{prop} \label{prop:value-iteration}
Value iteration (\cref{alg:value-iteration}) solves a DP problem
in $O(|S||A|T)$ queries to the oracle \eqref{eq:dp-classical-oracle}.
\end{prop}

\begin{alg}[h]
\noindent\framebox{\begin{minipage}[b]{0.986\textwidth}
\begin{algorithmic}[1]
\Procedure{ValueIteration}{$s_0$}
  \State Initialize an array $v[s]\gets 0$ for all $s \in S$
  \For{$t= T-1, T-2, \ldots, 1$}
    \For{$s \in S$}
      \State $w[s]\gets \max_{a} \left\{r_t(s, a) + v[a_t(s)]\right\}$
      \label{alg-line:query}
    \EndFor 
    \State $v\gets w$ 
  \EndFor
  \State \textbf{return} $\argmax_{a} \left\{r_0(s_0, a) + v[a_0(s_0)]\right\}$
\EndProcedure
\end{algorithmic}
\end{minipage}}
\caption{Value iteration}\label{alg:value-iteration}
\end{alg}

We let $S_t \subseteq S$ be the set of all states that are reachable at time
$t \in \mathbb T$. We call a DP problem \emph{time ordered} whenever the sets
$S_t$ form a partitioning of $S$, $S= \bigsqcup_{t \in \mathbb T} S_t$,
that is, a state $s \in S_t$ is only reachable at a time epoch $t$. We note that
a DP problem that is not readily time ordered can be turned into one that is by
replicating every state $s \in S$ to at most $O(T)$ copies $(s, t) \in
S \times \mathbb T$. Nevertheless, working with time-ordered DP problems allows
us to simplify the query complexity of solving them using Bellman's criteria via
the following algorithm.

\begin{prop} \label{prop:bellman-recursion}
Bellman's recursion (\cref{alg:bellman-recursion}) solves a
time-ordered DP problem in $O(|S||A|)$ queries to the oracle
\eqref{eq:dp-classical-oracle}.
\end{prop}

\begin{alg}
\noindent\framebox{\begin{minipage}[t]{0.986\textwidth}
\begin{algorithmic}[1]
\Procedure{BellmanRecursion}{$s_0$}
  \State Initialize a stack $R= \{(s= s_0, t=0)\}$
  \State Initialize an array $v[(s, t)]\gets \emptyset$
    for all $s \in S$ and $t \in \mathbb T \cup \{T\}$
  \State Assign $v[(s, T)]\gets 0$ for all $s \in S$
  \While{$R \neq \emptyset$}
    \State Get $(s, t)$ from $R$
    \If{$v[(a_t(s), t+1)]\neq \emptyset$ for all $a \in A$}
      \State $v[(s, t)]\gets \max_{a} \left\{r_t(s, a) + v[a_t(s), t+1]\right\}$
      \State Pop $(s, t)$ from R
    \Else
      \State Push $(a_t(s), t+1)$ to $R$ for all $a \in A$ such that
      $v[(a_t(s), t+1)]= \emptyset$
    \EndIf
  \EndWhile
  \State \textbf{return} $\argmax_{a} \left\{v[(a_0(s_0), 0)]\right\}$
\EndProcedure
\end{algorithmic}
\end{minipage}}
\caption{Bellman's recursion}\label{alg:bellman-recursion}
\end{alg}

\section{Open Problems for Quantum Computation}
\label{sec:open-quantum-problems}

As will be apparent from the examples presented in \cref
{sec:examples}, typically the number of \mbox{states $|S|$} is exponentially
larger than the time horizon $T$ and the number of actions $|A|$. Therefore,
providing quantum speedups in terms of $|S|$ is of particular interest to us,
and computational complexity factors of the form $\poly(|A|, T)$ are considered
negligible. The oracle above can be efficiently constructed
using $\plog(|S|)$ qubits and the same order of customary preliminary
gate sets (e.g., the Clifford+T set) in practical cases of interest.
In view of \cref{prop:value-iteration}, we now present the following problem.
\begin{prob}
\label{prob:quantum-dp}
Does there exist a bounded-error quantum algorithm that returns
the $s_0$ component of the solution to \eqref{eq:dp-problem} using
$\widetilde O(\sqrt{|S|}\poly(|A|, T))$ queries to the oracle
\eqref{eq:dp-coherent-oracle}?
\end{prob}
\noindent Achieving such a speedup for solving DP problems has appeared to be
a challenging open problem. See \cref{sec:related} for a summary of recent
attempts.

A quantum variant of the value iteration operator
\eqref{eq:classical-value-iteration} can be viewed as a unitary
transformation that receives a register prepared in the superposition of a
set of indices $s \in S_t$ and a set of associated values $v(s)$ in the
computational basis, and performs the transformation
\begin{equation}
\label{eq:q-val-iter}
U_t: \sum_{s \in S_t} \ket{s}\ket{v(s)}{\ket 0} \mapsto
\sum_{s \in S_t} \ket{s}\ket{v(s)}\ket{w(s)},
\end{equation}
where $w(s)= \max_a \left(r_t(s,a) + v(a_t(s))\right)$ for a given
$t \in \mathbb T$. Given \cref{alg:value-iteration}, if $U_t$ can be implemented
using $\widetilde O(\sqrt{|S_t|}\poly(|A|))$ queries to the oracle
\eqref{eq:dp-coherent-oracle}, then the answer to \cref{prob:quantum-dp} is
positive. More abstractly, we ask the following question.

\begin{prob}
\label{prob:q-val-iter}
Let $f: X \to \mathbb Z_{\geq 0}$ be an integer-valued function on a discrete
finite domain $X$. Let $Y= \{\sigma_i: X \to X\}$ be a finite set of mappings
from $X$ to itself. Does there exist a unitary transformation
\begin{equation}
\label{eq:q-val-iter-abstract}
U: \sum_{x \in X} \ket{x}\ket{f(x)}{\ket 0} \mapsto
\sum_{x \in X} \ket{x}\ket{f(x)}\ket{\max_\sigma (f(\sigma(x)))}
\end{equation}
that uses $\widetilde O(\sqrt{|X|} \poly(|Y|))$ queries to the oracle
$\mathcal O: \ket{x}\ket{\sigma}\ket{z} \mapsto
\ket{x}\ket{\sigma}\ket{z \oplus \sigma(x)}$?
\end{prob}

We note that the quadratic scaling in \cref{prob:q-val-iter} is with respect to
the size of the domain $X$ as opposed to the size of $Y$, over which the
optimization is performed. This is unlike the behaviour expected from amplitude
amplification. Value iteration implies that a scaling better than
$\widetilde O(\sqrt{|X|})$ in \cref{prob:q-val-iter} would contradict the lower
bounds proven in \cref{sec:q-lbd}. That is, an algorithm solving
\cref{prob:q-val-iter} will solve the following problem as well.

\begin{prob}
\label{prob:quantum-dp-weak}
Does there exist a bounded-error quantum algorithm that returns the
$s_0$ component of the solution to \eqref{eq:dp-problem} using
$\widetilde O(\sum_{t \in \mathbb T}\sqrt{|S_t|}\poly(|A|))$ queries to the
oracle~\eqref{eq:dp-coherent-oracle}?
\end{prob}

\noindent We note that for time-ordered DP problems
$\sum_{t \in \mathbb T}\sqrt{|S_t|} \geq \sqrt{|S|}$. Therefore, it is useful
to distinguish the claim of this problem from the following stronger claim.

\begin{prob}
\label{prob:quantum-dp-strong}
Does there exist a bounded-error quantum algorithm that returns the
$s_0$ component of the solution to \eqref{eq:dp-problem} for a time-ordered
DP problem using $\widetilde O(\sqrt{|S|}\poly(|A|))$ queries to the
oracle~\eqref{eq:dp-coherent-oracle}?
\end{prob}

A quantum algorithm solving \cref{prob:quantum-dp-weak} also solves
\cref{prob:quantum-dp}, and a quantum algorithm solving either of
\cref{prob:q-val-iter,prob:quantum-dp-strong} also solves
\cref{prob:quantum-dp-weak}.

\subsection{Related problems}
\label{sec:related}

\paragraph{Linear programming with high precision}

We can write a linear program (LP) that is equivalent to the functional
equation \eqref{eq:func-eq}. The value function depends on the time epochs
$t \in \{0, \ldots, T\}$ and states $s \in S$. For each value $V^*_t(s)$ of the
value function, we assign a real variable $v_{s, t}$ and, for consistency, write
the constants $r_t(s, a)$ as $r_{s, a, t}$. The linear programming formulation
is as follows:
\begin{equation}
\label{eq:dp-lp}
\begin{split}
\min &\quad v_{s_0, 0} \\
\text{s.t.} &\quad v_{s, t} \geq r_{s, a, t} + v_{a(s), t+1} \quad
\forall\, a \in A, s \in S, t \in \mathbb T \\
& \quad v_{s, t} \geq 0 \quad \forall s \in S, t \in \mathbb T \cup \{T\}
\end{split}
\end{equation}
It is easy to check that the above LP is feasible and attains a unique solution.
In this unique solution, $v_{s, T}= 0$ for all $s \in S$.

Intuitively, the LP can be thought of as the formulation of a network flow
problem wherein the inward flow of each node $(s, t)$ must match the largest
outward flow of it toward the states $(a(s), t+1)$ for all $a \in A$ with the
addition of a flow bias in the amount of $r_{s, a, t}$. We would like to find
the smallest required inward flow from the initial node $(s_0, 0)$.

In an earlier preprint \cite{ronagh2019quantum}, the author attempted to solve
this LP using the multiplicative weight update method
(MWUM). This technique was previously used in
\cite{brandao2017quantum,van2017quantum} to solve semidefinite and linear
programming problems. It turned out that the scaling of the method in the
precision parameter of the solution prohibits the providing of a quadratic
quantum advantage.
Ignoring other factors, the MWUM requires $O(1/\epsilon^2)$ queries to return an
$\epsilon$-feasible solution (a point that is $\epsilon$ away from the feasible
domain of the LP in the $L^1$ norm). This
scaling in precision is the main drawback of MWUM. In particular, the proof
of \cite[Theorem III.5]{ronagh2019quantum} cannot be reduced to the case of
basic feasible solutions of the LP. The fractional approximate solutions of the
LP can incur exponentially many small amounts of error and result in
the readout of a suboptimal solution. In the network flow analogy, this
amounts to $O(|S|)$ of the nodes of the graph incurring an
$\epsilon= O (1/|S|)$ deficit in the outward flow they are supposed to generate.
This adds up to an $O(1)$ error in the approximation of $v_{a(s_0), 1}$, which
is enough to disguise the optimal action at~$s_0$.

We note that, assuming $|A|$ and $T$ are polylogarithmic in $|S|$, the number of
variables $n$ and the number of constraints $m$ in the LP \eqref{eq:dp-lp} are
both $\widetilde O(|S|)$. In the context of MWUM, the primal \emph{width} $\ell$
of \eqref{eq:dp-lp} (i.e., a bound on the optimal value of the objective of the
LP) and its dual width $L$ (i.e., a bound on the slack of the constraints of
the LP) are both $O(T\ceil{r})$, where $\ceil{r}$ is an upper bound on the
reward structure, as introduced in \cref{sec:dp}.

For generic algorithms for solving LPs, the parameters $\ell$, $L$, and
$\frac{1}{\epsilon}$ are related such that for equivalent LPs the quantity
$\eta= \frac{\ell L}{\epsilon}$ is invariant. For \eqref{eq:dp-lp}, we have
$\eta= O(T^2 \ceil{r}^2 |S|)$. Therefore, for a generic LP solver
to provide a quadratic speedup in solving \eqref{eq:dp-lp}, a scaling of
$O(\sqrt{\max\{n, m\}}\plog(\eta))$ is required. However, \cite{van2017quantum}
shows that any generic quantum LP solver with sublinear dependence on $n$ or
$m$ has to depend at least polynomially on $\eta$. Therefore, the desired
$\plog(\eta)$ dependence is not possible.

\paragraph{Coherent computation of convex conjugates of functions}

Another attempt at solving DP problems using quantum computation is reported
in \cite{sutter2020legendre, sutter2020quantum}, wherein the authors' aim was
to demonstrate a quadratic quantum speedup for DP problems for which the value
functions are convex.

Let $f: D \to \mathbb R$ be a convex function defined on a bounded real domain.
The argument of \cite{sutter2020legendre} relies on the existence of a unitary
transformation that evolves a register prepared in the superposition of the
values of $f$ to the superposition of the values of the convex conjugate
$f^*: K \to \mathbb R$ of this function defined on a dual bounded domain $K$
via $f^*(s)= \sup_{x \in D} (\langle s, x\rangle - f(x))$.
Solving convex DP problems is thus reduced to the efficient implementation of
the evolution
$$\sum_{x \in D} \ket{x}\ket{f(x)} \mapsto
\sum_{y \in K} \ket{y}\ket{f^*(y)}$$
using $\plog(|D|, |K|)$ quantum gates. However, the existence of such a
unitary is an open problem. We note that such a transformation resembles the
evolution \eqref{eq:q-val-iter-abstract} in \cref{prob:q-val-iter}.

\section{Examples}
\label{sec:examples}

\subsection{The travelling salesperson problem}

Let $G$ be a fully connected graph with vertices $V= \{1, \ldots, n\}$. We let
$1$ be a fixed starting vertex and $c_{ij}$ be the cost of travelling from
vertex $i$ to vertex $j$. The goal is to find a Hamiltonian cycle (a cycle that
visits each vertex of the graph exactly once) starting and ending at $1$, while
incurring the lowest total cost. The best known classical algorithm for TSP is
due to Bellman \cite{bellman1962dynamic} and Held and Karp
\cite{held1962dynamic} (BHK), and performs DP with a runtime of $O(n^2 2^n)$.

We define a state to be a pair $(H, i)$, where $i \in H$ and $H \subseteq V$.
An action at a state $(H, i)$ corresponds to the choice of a vertex
$j \in H \setminus \{i\}$. The instantaneous
cost of travelling from state $(H, i)$
to $(H \setminus \{i\}, j)$ is the cost of travelling from vertex $j$ to $i$,
that is, $c_{ji}$. The cost function $C(H, i)$ represents the minimum total cost
of a Hamiltonian path starting at $1$, entering $H$ immediately, traversing
$H$, and ending at $i$. Bellman's optimality criteria may now be written as
$$C(H, i) = \min_{j \in H \setminus\{i\}}
\Big[C(H \setminus\{i\}, j) + c_{ji}\Big].$$

Note that it is trivial to move from a cost-minimizing formulation to a
reward-maximizing one by assigning $r_{ij} = \ceil{c} + 1 - c_{ij}$, where
$\ceil{c}$ is an upper bound on the edge weights $c_{ij}$. The definition
of states $(H, i)$ can be extended to allow $i \not\in H$ and the
definition of the action of $j$ on $(H, i)$ can be extended to allow
$j \not\in H \setminus \{i\}$. For every singleton $H= \{i\}$, any action $j$
maps $(\{i\}, i)$ to $(\emptyset, j)$ with reward $c_{1i}$. Otherwise, when
$j \not\in H \setminus \{i\}$ or if $i \not\in H$, then the action of $j$ maps
the state $(H, i)$ to the state $(H \setminus \{i\}, j)$ with reward $0$.
We may now rewrite the DP problem as the problem of solving
the functional equation
$$V^*(H, i) = \max_{j \in H \setminus \{i\}}
\Big[V^*(H \setminus\{i\}, j) + r_{ji}\Big]$$
with boundary condition $V^*(\emptyset, j)= 0$ for all $j$.

\begin{rmk} This DP problem is time ordered. It includes $|S|= O(n 2^n)$ states,
$|A|= O(n)$ actions, and a time horizon of $T= O(n)$. Therefore, the BHK
algorithm has a time complexity of $O(n^2 2^n)= O(|S||A|)$.
\end{rmk}

\paragraph{Oracle construction}

We begin by assuming an oracle $U_G$ for the adjacency matrix of the
edge-weighted graph $G$:
$$\ket{i}\ket{j}\ket{x} \mapsto
\ket{i}\ket{j}\ket{x \oplus c_{ji}}.$$
The registers in $U_G$ require $2\log(n) + \log(\ceil{c})$ qubits. By preparing
$O(n^2)$ registers in the values $c_{ij}$, we obtain an implementation of the
oracle $U_G$ using $O(n^2\plog(n, \ceil{c}))$ qubits. From $U_G$ we can
construct an oracle similar to \eqref{eq:dp-coherent-oracle}:
\begin{equation}
\label{eq:tsp-q-oracle}
U_{\text{TSP}}= \ket{H, i}\ket{j}\ket{x}\ket{y} \mapsto
\ket{H, i}\ket{j}\ket{x \oplus \left[H \setminus \{i\}, j\right]}
\ket{y \oplus r_{ij}}.
\end{equation}
Every state $\ket{H, i}= \ket{H}\ket{i}$ is encoded using a binary string of
size $n$ that represents the subset $H \subseteq V$ and an index $i$ encoded
using $\log(n)$ qubits. Therefore, the registers in $U_{\text{TSP}}$
are made from $O(n\plog(n, \ceil{c}))$ qubits. The circuit $U_{\text{TSP}}$
queries $U_G$ and thus uses a total of $O(n^2 \plog(n, \ceil{c}))$ qubits.

\begin{prop} The oracle \eqref{eq:tsp-q-oracle} can be constructed using
$O(n^2 \plog(n, \ceil{c}))$ qubits and a similar order of elementary quantum
gates.
\end{prop}

\begin{rmk} Problem \eqref{prob:quantum-dp}, therefore, asks whether there
exists a quantum algorithm for TSP that makes $O^*(\sqrt{2^n})$ queries to the
oracle \eqref{eq:tsp-q-oracle}. We note that
$$
\sum_{k=0}^n \sqrt{k\binom{n}{k}}
\leq n^{1/2} \sum_{k=0}^{n} \sqrt{\binom{n}{k}}
\leq n^{3/2} \sqrt{\binom{n}{n/2}}
= O\left(n^{3/2}\left(\frac{2}{n\pi}\right)^{1/4}2^{n/2}\right),
$$
where the equality follows from Stirling's approximation
$\binom{n}{n/2} \sim (\frac{2}{n\pi})^{1/2}2^n$. Therefore, answers in the
affirmative to \cref{prob:quantum-dp-weak,prob:quantum-dp-strong}
would provide similar quantum speedups.
\end{rmk}

\begin{rmk} \cite{ambainis2019quantum} shows a bounded-error quantum algorithm
for solving TSP that uses recursive applications of Grover’s search to solve
this problem in $O^* (1.728^n)$. However, this algorithm requires QRAM access
to the classical BHK algorithm on graphs of size $0.24n$, run in superposition.
\end{rmk}

\subsection{The minimum set-cover problem}
\label{sec:msc}

Consider a set $U$, called the \emph{universe}, with $n$ elements, and
a family $\mathcal F= \{V_1, \ldots, V_m\}$ of $m$ subsets
$V_i \subseteq U$. The minimum set-cover problem (MSC) is the problem of
finding the minimum number of these subsets required to cover the entire
universe. That is, the goal is to find the minimum cardinality
$F \subseteq \mathcal F$ such that $ \bigcup_{V \in F} V = U$.
We will use the notation $\bar F$ to denote the union of all elements of the
members of $F$, so $\bar F = \bigcup_{V \in F} V$.

We define a DP problem as follows. The states are the subsets
$S \subseteq U$ of the universe. There are only two actions $A= \{u, v\}$
where the transition from $S$ via $u$ at time $t$ is the inclusion of $V_{t+1}$
and the transition via $v$ skips this inclusion. Hence,
$$u_t: S \mapsto S \cup V_{t+1}\quad \forall S \subseteq U
\qquad \text{and} \qquad
v_t: S \mapsto S\quad \forall S \subseteq U.$$
The transition via $v$ occurs at no additional cost, whereas transition via
$u$ adds a new set to the candidate set cover. To remain in a
reward-maximizing framework, we therefore define the reward for transition via
$v$ as $1$ and the reward for transition via $u$ as $0$. The actions
$u_n$ and $v_n$ send any state $S$ to itself with a reward of $0$ if $S \neq U$
and a reward of $m + 1$ for the state $U$.
We mark an initial state $s_0 = \emptyset$ at time $t = 0$. It is
straightforward to see that the value function at $s_0$ is maximized by a
policy that constructs a minimum set cover.

\begin{rmk} The DP problem has a time horizon $T= O(m)$, $|A|= 2= O(1)$ actions,
and $|S|= O(2^n)$ states. The best known classical algorithm for MSC is the
above DP solution \cite{fomin2013exact}. The runtime is $O(nm2^n)$, consisting
of $O(m 2^n)= O(T|S|)$ queries to the classical oracle
\eqref{eq:dp-classical-oracle} and the oracle itself contributing an additional
$O(n)$ factor (for set operations).
\end{rmk}

\begin{rmk} This DP problem is not time ordered. However, one can replace the
definition of states from subsets $S \subseteq U$ to pairs $(S, k)$ of a subset
$S \subseteq U$ and an integer $k=0, \ldots, m$. Then, the DP problem becomes
time ordered with $|S|= O(m 2^n)$.
\end{rmk}

The family $\mathcal F$ can be prepared using $O(mn)$ qubits by encoding any
set $V_i \subseteq U$ using a binary string of size $n$. Forming unions and set
comparisons can be done using $O(n)$ elementary quantum gates. This suffices
for efficient construction of an oracle
\begin{equation}
\label{eq:msc-q-oracle}
U_{\text{MSC}}:
\ket{S}\ket{t}\ket{a}\ket{x}\ket{y} \mapsto
\ket{S}\ket{t}\ket{a}\ket{x \oplus a_t(S)}
\ket{y \oplus r_{s, a, t}},
\end{equation}
where $a\in \{u, v\}$.

\begin{prop} The oracle $U_{\mathrm{MSC}}$ can be constructed using $O(mn)$
qubits and the same order of elementary gate operations.
\end{prop}

\begin{rmk} \cref{prob:quantum-dp} asks
whether MSC can be solved in $O^*(\sqrt{2^n} \poly(m))$ queries to
the oracle~\eqref{eq:msc-q-oracle}, while
\cref{prob:quantum-dp-weak,prob:quantum-dp-strong} ask for
query complexities $O^*(m\sqrt{2^n})$ and $O^*(\sqrt{m2^n})$, respectively.
\end{rmk}

\begin{rmk} \cite{ambainis2019quantum} shows a bounded-error quantum algorithm
for solving MSC that uses recursive applications of Grover’s search to solve
this problem in $O (1.728^n\poly(m, n))$ using QRAM.
\end{rmk}

\subsection{The edit distance problem}

Given two strings $x$ and $y$, find the small sequence of edit operations that
will transform $x$ to $y$. The edit operations consist of substitution of one
character for another, the removal of a character, and the insertion of a new
character. Let $n= |x|$ and $m= |y|$ be the original sizes of the strings. We
define a state $s(i, j)$ for all $i \in \{1, \dots, n\}$ and
$j \in \{1, \dots, m\}$. Each state represents the pair of strings
$(x[1:i], y[1:j])$. So, the initial state is $(n, m)$ and there
are three actions, $A= \{\rho, \iota, \delta\}$, acting via
$$\rho: (i, j) \mapsto (i-1, j), \quad
\iota: (i, j) \mapsto (i, j-1), \quad
\delta: (i, j) \mapsto (i-1, j-1),$$
respectively representing the removal of the last character from $x'$, the
insertion of the last character in $y'$, and the change of the last character
of $x'$ to the last character of $y'$.
The cost of these actions is time homogeneous and is defined as $1$ for
$\rho$, $1$ for $\iota$, and $c=0$ for $\delta$ when the last characters of
$x'$ and $y'$ are the same, and $c=2$ when the last characters are different.
We will switch around the costs $0$ and $2$ to achieve a reward of $r= 2 - c$
for these actions, and a reward-maximizing formulation for the DP problem:
$$r((i, j), \rho, t)= 2\delta(x[i], y[j]) \quad \forall t \in \mathbb T,$$
where $\delta$ is the Kronecker delta on the set of characters.

Bellman's recursion is therefore written as
$$V^*(i, j)= \max(
V^*(i-1, j)+ 1,
V^*(i, j-1)+ 1,
V^*(i-1, j-1)+ 2\delta(x[i], y[j])),$$
with the boundary conditions $V^*(i, 0)=i$ and $V^*(0, j)= j$. Alternatively,
we can extend the definitions of the actions and their rewards by
\begin{align*}
\rho: (i, j) &\mapsto (\max(0, i-1), j),
\quad\text{ with reward } 1- \delta(i, 0); \\
\iota: (i, j) &\mapsto (i, \max(0, j-1)),
\quad\text{ with reward } 1- \delta(j, 0); \text{ and} \\
\delta: (i, j) &\mapsto (\max(0, i-1), \max(0, j-1)),
\quad\text{ with reward } 2\delta(x[i], y[j]) - \delta(i, 0) + \delta(j, 0).
\end{align*}
We have $|S|= O(nm)$, $|A|= O(1)$, and $T= O(n+m)$. The transition
kernel and reward structure are both time homogeneous; therefore, Bellman's
recursion succeeds in $O(|S|)$ queries \cite{wagner1974string}.

When $n= m$, the DP runtime is $O(n^2)$ and, under the strong exponential time
hypothesis, the problem cannot be solved in a time of $O(n^{2- \epsilon})$.
We also note that, for $m < n$, \cite{ambainis2020quantum} provides a
quantum query complexity lower bound of $\Omega((\sqrt{n}m)^{1-\epsilon})$.

\begin{rmk}
We can achieve a time-ordered formulation by replicating each state at all
points $t$ in time from which that state is accessible. A state $(i, j)$ is
reachable in the window of time \mbox{$t= \max(i, j), \ldots, i+j$}. Therefore,
the cardinality of the set
$$S_t= \{s: s \text{ is accessible at time } t\}
= \{ (i, j): i, j \leq t, i+j \geq t\}$$
is $O(t^2)$. We note that
$$\sum_{t=0}^{n+m} \sqrt{|S_t|}=
2 \sum_{t=1}^m \sqrt{\frac{t^2}2} + (n-m)\sqrt{\frac{m^2}2}=
O(m^2) + O((n-m)m)= O(nm),
$$
whereas
$$\sqrt{\left(\sum_{t=0}^{n+m} |S_t|\right)}=
\sqrt{2 \sum_{t=1}^m \frac{t^2}2 + (n-m)\frac{m^2}2}=
\sqrt{O(m^3) + O((n-m)m^2)}= O(\sqrt{n}m).
$$
Therefore, \cref{prob:quantum-dp-weak} does not provide a quantum advantage
but an answer in the affirmative to \cref{prob:quantum-dp-strong} would close
the gap with the lower bound provided in \cite{ambainis2020quantum}.
\end{rmk}

\section{Quantum Complexity Lower Bound}
\label{sec:q-lbd}

\begin{figure}[b]
\scalebox{.8}{\begin{tikzpicture}
[scale=2, thick,
nodeDecorate/.style={shape=circle,fill=gray!10,inner sep=3pt,draw,thick}]

\foreach \nodename/\x/\y/\label in {
  s00/0/0/s_0,
  s01/1/-1/-,
  s02/1/0.75/-,
  s03/2/-1.5/-,
  s04/2/-0.75/-,
  s05/2/0.25/-,
  s06/2/1.25/-}
{
  \node (\nodename) at (\x,\y) [nodeDecorate]
  {\scriptsize \ifthenelse{\equal{\label}{-}}{$\quad\!$}{$\label$}};
}

\foreach \i in {1,...,7}{
  \node (s1\i) at (3.5,{0.5*\i-2)}) [nodeDecorate]
  {\ifthenelse{\equal{\i}{2}}{$\bf \bar s$}{$\quad\!$}};
}

\tikzstyle{LabelStyle}=[fill= white]
\foreach \i in {1,...,7}{
  \node (sB\i) at (5,{0.5*\i-2)}) [nodeDecorate] {$\quad\!$};
  \scriptsize
  \Loop[label= 0, dist= 0.5cm, dir= EA, style= {->}](sB\i)
}

\foreach \i in {1}{
  \node (sG\i) at (6.5,{0.5*\i-2)}) [nodeDecorate] {$\quad\!$};
  \scriptsize
  \Loop[label= 2, dist= 0.5cm, dir= EA, style= {->}](sG\i)
}

\tikzstyle{EdgeStyle}=[->, thick]
\tikzstyle{LabelStyle}=[fill= white]

\foreach \startnode/\endnode/\bend/\action/\ldist/\langle/\reward in {
  s00/s01/bend left=0/a_{\text{R}}/-1mm/180/0,
  s00/s02/bend left=0/a_{\text{L}}/-1mm/180/0,
  s01/s03/bend left=0/a_{\text{R}}/0mm/180/0,
  s01/s04/bend left=0/a_{\text{L}}/0mm/175/0,
  s02/s05/bend left=0/a_{\text{R}}/0mm/180/0,
  s02/s06/bend left=0/a_{\text{L}}/0mm/180/0,
  s03/s11/bend left=30/a_{\text{L}}/2.5mm/180/0,
  s03/s11/bend left=-30/a_{\text{R}}/2.5mm/180/0,
  s04/s12/bend left=0/a_{\text{R}}/2.5mm/180/0,
  s04/s13/bend left=0/a_{\text{L}}/2.5mm/180/0,
  s05/s14/bend left=0/a_{\text{R}}/2.5mm/180/0,
  s05/s15/bend left=0/a_{\text{L}}/2.5mm/180/0,
  s06/s16/bend left=0/a_{\text{R}}/2.5mm/180/0,
  s06/s17/bend left=0/a_{\text{L}}/2.5mm/180/0
} {
  {\scriptsize\tikzset{LabelStyle/.style=
  {label={[label distance=\ldist]\langle:$\action\,$}}}
  \Edge[label= $\reward$, style= \bend](\startnode)(\endnode)}
}

\draw[color=gray!80] (-0.35,0) to[out=50,in=-180]
      (2.25,1.75) --
      (2.25,-2) to[out=180,in=-50] (-0.35,0);
\node () at (0.25,1) {\Large $S_\top$};

\draw[color=gray!80] (3.1,-2) rectangle (4,2);
\node () at (2.95,1.8) {\Large $S_1$};

\draw[color=gray!80] (4.6,-2) rectangle (5.5,2);
\node () at (4.43,1.8) {\Large $S_2$};

\node () at (6.25,-1.25) {\Large $S_\bot$};

\foreach \startnode/\endnode in {
  s11/sB2,
  s11/sB3,
  s11/sB6,
  s12/sB2,
  s12/sB3,
  s13/sB3,
  s13/sB4,
  s13/sB7,
  s14/sB5,
  s14/sB1,
  s14/sB4,
  s15/sB6,
  s15/sB5,
  s15/sB4,
  s16/sB7,
  s16/sB6,
  s16/sB4,
  s17/sB1,
  s17/sB3,
  s17/sB4}
{
  \Edge[label={\scriptsize 0}](\startnode)(\endnode)
}

\tikzset{LabelStyle/.style={label=right:$\bf \bar a$}}
\Edge[label= {\scriptsize $0$}, style= {dashed}](s12)(sB1)

\tikzset{LabelStyle/.style={label=right:$\bf \bar a$}}
\Edge[label= {\scriptsize $2$},
style= {dashed, bend right=40}](s12)(sG1)

\end{tikzpicture}}
\caption{Schematics of instances in $\mathcal M_1$ and $\mathcal M_2$. A pair
$M_1 \in \mathcal M_1$ and $M_2 \in \mathcal M_2$ of DP instances is depicted
that are in relation $R$, as their transition kernels differ in a single
state--action pair $(\bar s , \bar a) \in S_1 \times A$.}
\label{fig:c-lbd}
\end{figure}
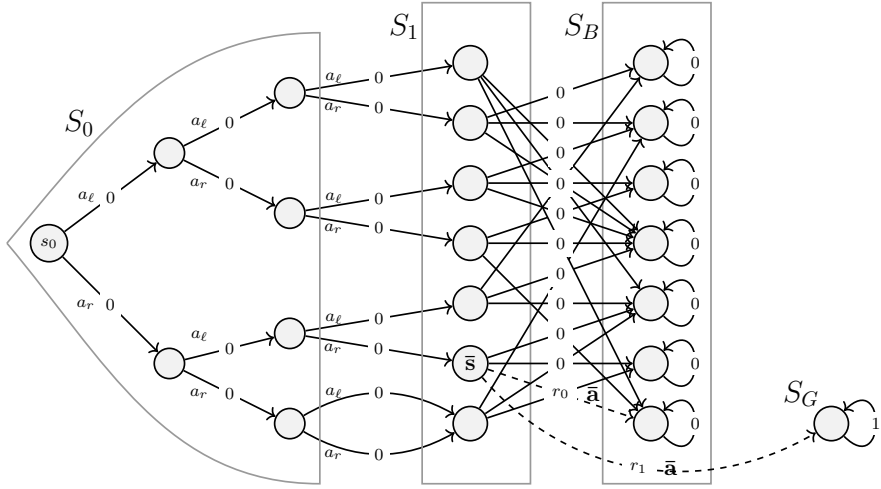

We now investigate the quantum query complexity of solving DP problems using the
adversary method of \cite{ambainis2002quantum}. Our construction follows ideas
from \cite{chen2017lower}. Consider two families of DP problem instances
$\mathcal M_1$ and $\mathcal M_2$, depicted in \cref{fig:c-lbd}. The two families
share the same state space \mbox{$S = S_\top\sqcup S_1\sqcup S_2\sqcup S_\bot$,}
the same action space $A$, and the same time horizon $T \in \mathbb N$. We let
$S_1= S_2= [n]= \{1, \ldots, n\}$ and assume that $|A|> 2$. The set $S_\bot$ is a
singleton $|S_\bot|=1$. For all instances in $\mathcal M_1$ and $\mathcal M_2$,
every action maps $s \in S_\bot$ to itself with a reward of $2$ and every
$s \in S_2$ to itself with a reward of $0$.

The structure of $S_\top$ is also common between DP problem instances in
$\mathcal M_1$ and $\mathcal M_2$. It contains the initial state $s_0 \in
S_\top$. Let $a_{\text{L}}, a_{\text{R}} \in A$ be two fixed actions. The
states in $S_\top$ form a binary tree with $s_0$ as the root. The role of
$S_\top$ is to make every state in $S_1$ accessible from $s_0$ in
$\ceil {\log n}$ steps. The actions $a_{\text{L}}$ and $a_{\text{R}}$ map
every parent state to its left and right children (which might coincide) with a
reward of $0$, and every action
$a \in A \setminus \{a_{\text{L}}, a_{\text{R}}\}$ maps
every state in $S_\top$ to itself with a reward of $1$. It is easy to see that
$|S_\top| \leq 2n$ and thus $|S| = O(n)$.

For any $M_1 \in \mathcal M_1$, every $a \in A$ maps every $s \in S_1$ to some
$a(s) \in S_2$ with a reward of $0$. Therefore, the optimal value function for
$M_1$ at $s_0$ is $v_{M_1}^*(s_0)= T$ and any action
$a \neq a_{\text{L}}, a_{\text{R}}$ is optimal. The instances
$M_2 \in \mathcal M_2$ differ from those in $\mathcal M_1$ only in a special
state--action pair $(\bar s, \bar a) \in S_1 \times A$ for which
$\bar a(\bar s)$ is the single element of $S_\bot$ with a reward of $\bar r= 2$.
So long as $T > 2\ceil{\log(n)}$, the optimal action at $s_0$ is one of
$a_{\text{L}}$ and $a_{\text{R}}$, depending on the choice of $\bar s$. We note
that in the argument that follows we could instead assume
$T > \ceil{\log(n)}$ but use $\bar r = T$. However, this would impose a scaling
constraint of $\ceil{r} = \Omega(\log(n))$ on the reward structure.

Now, consider a function $f: \{0, 1\}^* \to \{0, 1\}$ that receives a binary
string describing the transition kernel of a problem instance in $\mathcal
M_1 \sqcup \mathcal M_2$ and returns $0$ if and only if the optimal action at
$s_0$ is in $\{a_{\text{L}}, a_{\text{R}}\}$.

\begin{thm}
\label{thm:q-adv-bound}
Any quantum algorithm that computes the function $f$ above uses
$\Omega(\sqrt{|S||A|})$ queries.
\end{thm}

\begin{proof}
We consider the relation $R$ between instances $M_1 \in\mathcal M_1$ and $M_2
\in \mathcal M_2$ to be defined as $(M_1, M_2) \in R$ if and only if their
transition kernel differs in exactly a single pair $(\bar s, \bar a)$. We now
use \cite[Theorem 2]{ambainis2002quantum}. We note the following:
\begin{itemize}\setlength\itemsep{1px}
\item Each instance in $\mathcal M_1$ is in relation $R$ with $|S_1||A|$
instances in $\mathcal M_2$;

\item Each instance in $\mathcal M_2$ is in relation $R$ with $|S_2|$ instances
in $\mathcal M_1$;

\item For every instance in $\mathcal M_1$ and every pair
$(s, a) \in S \times A$ there is at most $1$ instance in $\mathcal M_2$ with
a different transition kernel $(s, a) \mapsto \left(a(s), r(s, a)\right)$; and

\item For every instance in $\mathcal M_2$ and every pair
$(s, a) \in S \times A$ there are at most $|S_2|$ instances in $\mathcal M_1$
with a different transition kernel $(s, a) \mapsto \left(a(s), r(s, a)\right)$.
\end{itemize}
Then, \cite[Theorem 2]{ambainis2002quantum} implies that the number of queries
made by the quantum algorithm is lower bounded by
$$\Omega\left(\sqrt{\frac{|S_1||A||S_2|}{|S_2|}}\right)
= \Omega\left(\sqrt{|S_1||A|}\right)= \Omega(\sqrt{|S||A|})\,,$$
proving the theorem.
\end{proof}

\begin{cor}
\label{cor:lower-bound-dp}
A bounded-error quantum algorithm solving finite-horizon DP problems with
states~$S$, actions $A$, and time horizon $T= \Omega(\log(|S|))$ makes
$\Omega(\sqrt{|S||A|})$ queries to the oracle \eqref{eq:dp-coherent-oracle}.
\end{cor}

\begin{cor}
\label{cor:optimal-dp}
A bounded-error quantum algorithm solving \cref{prob:quantum-dp} is optimal
in $|S|$ for DP problems with $|A|= O(\plog(|S|))$ and time horizon
$T= \Theta(\plog(|S|))$.
\end{cor}

\begin{prop}
\label{prop:lower-bound-odp}
A bounded-error quantum algorithm solving time-ordered finite-horizon DP
problems with states $S$, actions $A$, and time horizon $T= \Omega(\log(|S|))$
makes $\Omega(\sqrt{|S||A|})$ queries to the oracle
\eqref{eq:dp-coherent-oracle}.
\end{prop}

\begin{proof}
We note that the DP families $\mathcal M_1$ and $\mathcal M_2$ are not time
ordered since the actions $a \neq a_{\text{L}}, a_{\text{R}}$ map the states
in $S_\top$ to themselves. However, this can be rectified by the addition of
$O(\log(n))$ states $\{u_t: t= 1, \ldots, \ceil{\log (n)}\}$ to $S_\top$. The
role of state $u_t$ is to ``absorb'' the actions of
$a \neq a_{\text{L}}, a_{\text{R}}$ at time $t-1$ from all states
in $S_\top$. It is easy to see that this modification turns $\mathcal M_1$
and $\mathcal M_2$ into time-ordered DP problems while the argument of
\cref{thm:q-adv-bound} remains valid.
\end{proof}

\begin{cor}
\label{cor:optimal-ordered-dp}
A bounded-error quantum algorithm solving \cref{prob:quantum-dp-strong}
for time-ordered finite-horizon DP problems with time horizon
$T= \Theta(\plog(|S|))$ is optimal in $|S|$, and dependence on a
$\poly(\sqrt{|A|})$ factor is inevitable.
\end{cor}

\section{Classical Complexity Lower Bound}
\label{sec:c-lbd}

We now investigate the computational complexity of solving DP problems
classically in an analogous but classical oracle setting.
Once again, we borrow techniques from adversary methods
\cite{aaronson2006lower,ambainis2002quantum,chen2017lower}, but this time apply
them to bounded-error classical randomized algorithms. As in \cref{sec:q-lbd},
we define families of DP instances $\mathcal M_1$ and $\mathcal M_2$ that share
the same state and action spaces. We then show that, if a randomized algorithm
solves DP problems with high probability, there should be a deterministic
algorithm $\mu$ that also succeeds in distinguishing a large fraction of the
instances in the two families.

The family $\mathcal M= \mathcal M_1 \sqcup \mathcal M_2$ of DP instances is
defined as in \cref{sec:q-lbd} and \cref{fig:c-lbd}. By a similar argument to
that in the previous section, it is obvious that an algorithm that finds an
optimal action at $s_0$ is able to distinguish instances between $\mathcal M_1$
and $\mathcal M_2$. Let $m= n|A|= |S_1||A|$. It is straightforward to see that
$|\mathcal M_1| = |\mathcal M_2| = m n^m$.

Let $\Pi_Q$ be the set of all the deterministic algorithms which, for an
instance $M \in \mathcal M$, make at most $Q$ queries to the oracle
\eqref{eq:dp-classical-oracle} given by $(s, a, t) \mapsto (a_t(s), r_t(s, a)$
before returning an optimal action at $s_0$.
A randomized algorithm running at most $Q$ steps is a distribution $\mu$ on
$\Pi_Q$. Let $\mathcal P(\Pi_Q)$ be the set of all probability measures on
$\Pi_Q$ and $a^\mu_M$ be the action returned by $\mu$ on input~$M$.
Suppose there exists a randomized algorithm $\mu \in \mathcal P(\Pi_Q)$ that,
when run on every $M \in \mathcal M$, correctly returns an optimal action
$a^\mu_M \in \pi^*_M(s_0)$ with high probability. That is to say,
\begin{equation}
\label{eq:thm}
\max_{\mu\in \mathcal P(\Pi_Q)}
\min_{M\in \mathcal M}
P_{a\sim\mu(M)} \Big(a^\mu_M \in \pi^*_M(s_0)\Big) \geq 1 - \xi\,,
\end{equation}
which by Yao's minimax principle implies
\begin{equation}\label{eq:yao}
\min_{D\in \mathcal P(\mathcal M)}
\max_{\mu \in \Pi_Q}
P_{M\sim D} \Big(a^\mu_M \in \pi^*_M(s_0)\Big) \geq 1 - \xi\,,
\end{equation}
where $D$ is a distribution on $\mathcal M$.

Let $D_1$ and $D_2$ be uniform distributions on $\mathcal M_1$ and
$\mathcal M_2$, respectively, and let $D$ be the uniform mixture of the two.
Now let $\mu \in \Pi_Q$ be a deterministic
algorithm which fails to return an optimal $a^\mu_M \in \pi^*_M(s_0)$ with a
probability of at most $\xi$ on inputs from $D$. This implies that $\mu$ fails
with a probability of at most $2 \xi$ if the instance is drawn from either of $D_1$
or $D_2$ considered individually. We define $\mathcal C_i\subset \mathcal M_i$
as the sets of instances for which $\mu$ succeeds. It is obvious that
$$|\mathcal C_i| \geq (1- 2\xi) |\mathcal M_i| = (1- 2\xi) mn^m\,.$$

We call $M_1 \in \mathcal M_1$ and $M_2 \in \mathcal M_2$ a \emph{twin} if
their transition kernels are identical except that the reward for taking action
$\bar a$ at state $\bar s$ is $r_i$ for $i= 1, 2$.
We let $E(\mathcal A_1, \mathcal A_2)$ denote the number of
twins where the $i$-th component of the twin is in $\mathcal A_i$ for
$i= 1, 2$. The number of twins on which $\mu$ succeeds is lower bounded by
\begin{align*}
E(\mathcal C_1, \mathcal C_2)
\geq E(\mathcal C_1, \mathcal M_2)
- E(\mathcal M_1, \mathcal M_2 \setminus \mathcal C_2)
\geq (1 - 2 \xi) mn^m - 2\xi m n^m = (1 - 4\xi) m n^m.
\end{align*}
Setting $\xi= \frac{1}{8}$ guarantees that $\mu$ distinguishes at least
$\frac{1}2 mn^m$ twins of the DP instances. The key observation now is that,
for any twin, $\mu$ has to query $(\bar s, \bar a)$, that is, the special
state--action pair associated to the twin; otherwise, $\mu$ cannot distinguish
them. We now define a new problem.

\begin{defn}[Function distinction]
\label{def:func-diff}
Let $f, g: X \to \{0, \ldots, n\}$ be two integer-valued functions defined on a
discrete domain $X= \{1, \ldots, m\}$. We say that a deterministic algorithm is
able to distinguish $f$ from $g$ if it queries a \emph{witness} point $x \in X$
for which $f(x) \neq g(x)$.
\end{defn}

We say $f$ and $g$ as given in the above definition form a twin if $f$ takes
only nonzero values and $g$ differs from $f$ in exactly one point $x \in X$ at
which $g(x)= 0$. We note that each DP instance in the families $\mathcal M_1$
and $\mathcal M_2$ corresponds uniquely to a function
$S_1 \times A \to S_2 \sqcup S_\bot$ and therefore to a function
$X \to \{0, \ldots, n\}$. Therefore, an algorithm $\mu$ as given above that
distinguishes twins of DP instances is equivalent to an algorithm that
distinguishes twins of functions.

\begin{prop}
\label{prop:matrix-diff-lb}
Any deterministic algorithm $\mu$ that performs vector differentiation needs
$\Omega(m)$ queries to distinguish at least $\frac{1}2 m n^m$ twins of
functions.
\end{prop}

\begin{proof}
We view the queries of $\mu$ as a decision tree. At every node of the tree,
$\mu$ queries its input function at a certain point in the domain. The root of
the tree is the beginning of the algorithm at which no queries have yet been
made. We say this node is at depth $0$. A node at which a $k$-th query to the
vector is made is called a depth-$k$ node. It is obvious that a depth-$k$ node
can distinguish at most $n^{m-k}$ pairs of functions. Let $(f, g)$ be a twin,
with $f$ and $g$ distinguishable at a depth-$k$ node. This means that all
previous $k-1$ queries to $f$ and $g$ have returned the same integers. The
$k$-th query has resulted in a nonzero integer for one of the functions and $0$
for the other. There are $m-k$ remaining entries and $f$ and $g$ have to
coincide for all of them. This means that there are $n^{m-k}$ ways to complete
$f$ and $g$ into twins.

On the other hand, there are at most $n^k$ nodes at a depth of $k$. Therefore,
the depth-$k$ nodes can in total distinguish at most $n^m$ twins of functions.
In order for $\mu$ to distinguish $\frac{1}2 m n^m$ twin functions, the total
depth of the decision tree of $\mu$ has to be at least $\frac{1}2m$. This
proves the claim.
\end{proof}

\begin{cor} Any classical randomized algorithm that solves a DP
problem at a marked initial state and a time horizon $T= \Omega(\log(|S|))$ via
queries to the oracle \eqref{eq:dp-classical-oracle} has to make at least
$\Omega(|S||A|)$ queries to that oracle.
\end{cor}

\section{Acknowledgement}

The author thanks Ronald~de~Wolf, Artur~Scherer, Seyed~Saeed~Changiz~Rezaei,
Yichen~Chen, Ryuhei~Mori, Yoichi~Iwata, Jevg\={e}nijs~Vihrovs,
Kri\v{s}j\={a}nis~Pr\={u}sis, J\={a}nis~Iraids, Martins~Kokainis,
and Scott~Aaronson for useful technical discussions. The author further thanks
Marko~Bucyk for his careful review and editing of this manuscript. The author
acknowledges the support of 1QBit, the Government of Ontario, and Innovation,
Science and Economic Development Canada.

\bibliography{bib}
\end{document}